# *Preliminary Report: Missense mutations in the APOL gene family are associated with end stage kidney disease risk previously attributed to the MYH9 gene*

**Shay Tzur[1,8], Saharon Rosset[2,8], Revital Shemer[1], Guennady Yudkovsky[1], Sara Selig[1,3], Ayele Tarekegn[4,5], Endashaw Bekele[5], Neil Bradman[4], Walter G Wasser[6], Doron M Behar[3,7], Karl Skorecki[1,3 *]**

[1]Ruth and Bruce Rappaport Faculty of Medicine and Research Institute, Technion - Israel Institute of Technology, Haifa 31096, Israel. [2]Department of Statistics and Operations Research, Tel Aviv University, Tel Aviv 69978, Israel. [3] Molecular Medicine Laboratory, Rambam Health Care Campus, Haifa 31096, Israel. [4]The Centre for Genetic Anthropology, Research Department of Genetics, Evolution and Environment, University College London, London WC1E 6BT, UK. [5]Department of Biology, Addis Ababa University, Addis Ababa 1176, Ethiopia. [6]Hadassah Medical Center, Jerusalem 91120, Israel. [7]Estonian Biocentre and Department of Evolutionary Biology, University of Tartu, Tartu 51010, Estonia. [8]These two authors contributed equally to this work.

*Correspondence: Karl Skorecki, 8 Ha'Aliyah Street, Haifa 35254, Israel. Tel: 972-4-8543250 fax: 972-4-8542333, email: skorecki@tx.technion.ac.il

**ABSTRACT**

**MYH9 has been proposed as a major genetic risk locus for a spectrum of non-diabetic end stage kidney disease (ESKD). We use recently released sequences from the 1000 Genomes Project to identify two western African specific missense mutations (S342G and I384M) in the neighbouring APOL1 gene, and demonstrate that these are more strongly associated with ESKD than previously reported MYH9 variants. We also show that the distribution of these risk variants in African populations is consistent with the pattern of African ancestry ESKD risk previously attributed to the MYH9 gene. Additional associations were also found among other members of the APOL gene family, and we propose that ESKD risk is caused by western African variants in members of the APOL gene family, which evolved to confer protection against pathogens, such as Trypanosoma.**

## INTRODUCTION

Mapping by admixture linkage disequilibrium (MALD) localized an interval on chromosome 22, in a region that includes the MYH9 gene, which was shown to contain African ancestry risk variants associated with certain forms of ESKD (Kao et al. 2008; Kopp et al. 2008). This led to the new designation of MYH9 associated nephropathies (Bostrom and Freedman 2010). Subsequent studies identified clusters of single nucleotide polymorphisms (SNPs) within MYH9 with the largest odds ratios (OR) reported to date for the association of common variants with common disease risk (Bostrom and Freedman 2010; Nelson et al. 2010). These MYH9 association studies were then extended to earlier stage and related kidney disease phenotypes, and to population groups with varying degrees of recent African ancestry admixture (Behar et al. 2010; Nelson et al. 2010). However, despite intensive efforts including re-sequencing of the MYH9 gene, no suggested functional mutation has been identified (Nelson et al. 2010). This led us to re-examine the MALD interval surrounding MYH9, and to the detection of novel missense mutations with predicted functional effects in the neighbouring APOL1 gene, that are significantly more associated with ESKD than all previously reported SNPs in MYH9. Additional associations were found among other members of the APOL gene family, and we propose that variants in members of the APOL gene family, which evolved to confer protection against pathogens, currently account for increased kidney disease risk in African ancestry populations, and account for the MALD peak previously attributed to MYH9.

## METHODS

### Sample sets

The ESKD sample set is a sub-set of a larger ESKD cohort previously reported, excluding individuals designated diabetic nephropathy as the cause of ESKD (Behar et al. 2010). The designation of MYH9 associated nephropathies included hypertension affiliated chronic kidney disease, HIV-associated nephropathy (HIVAN), and non-monogenic forms of focal segmental glomerulosclerosis (Bostrom and Freedman 2010). The sample includes 430 non-diabetic ESKD cases and 525 controls, all of whom are self-identified as either African American or Hispanic American. We have already reported the African, European and Native American ancestry admixture proportions for these samples, based on a set of 40 genome wide Ancestry Informative Markers, and also the association and risk parameters for a set of 42 SNP polymorphic markers in the MYH9 genetic locus in the entire sample set (Behar et al. 2010).

The African populations sample set consists of 676 samples from 12 African populations, including Cameroon (2 ethnic groups), Congo, Ethiopia (4 ethnic groups), Ghana (2 groups), Malawi, Mozambique, Sudan with details provided in Table 1. Whole Genome Amplification of extracted DNA was carried out as previously reported (van Eijk et al. 2010).

All sampling and testing was conducted with institutional review board approval for human genetic studies on anonymized samples, with informed consent.

## SNP Genotyping

Genotyping of the novel six SNPs reported herein was performed with the KasPar methodology (Petkov et al. 2004). In addition we also designed PCR and RFLP reactions for the APOL1 SNPs as described in the legend to Figure 1. SNP validation was performed by Sanger sequencing.

## Statistical Analysis

To search for variants possibly associated with ESKD, we examined a 1.55 Mbp interval surrounding MYH9, spanning nucleotide positions 34,000,000 to 35,550,000 (NCBI36 assembly) in 119 whole genome sequences recently released in the 1000Genomes Project (www.1000genomes.org). Of these, 60 are of European origin (HapMap CEU cohort), and 59 are of West-African origin (HapMap YRI cohort) yielding a total of 7,479 SNPs. We selected from these for further examination all SNPs which complied with the following conditions:

1. Minor allele frequency in the CEU cohort not exceeding 7.5% (9/120 chromosomes). This minor allele was designated as a putative "risk state" for this SNP.
2. Risk state in the YRI cohort at allele frequency exceeding 17.5% (21/118 chromosomes).
3. A minimal level of LD with the MYH9 S-1 SNP rs5750250. The criterion used was a chi-square test p-value not exceeding 0.15 for the 3*3 genotype table comparing each candidate SNP to the S-1 SNP in the YRI cohort.

All candidates which passed these requirements (total 250) were inspected in order to identify non-synonymous exonic SNPs, indicating a possible functional role. These non-synonymous SNPs were again examined for consistency with association patterns in the leading MYH9 intronic risk variants S-1 (rs5750250) and F-1 (rs11912763) (Nelson et al. 2010). The high attributable risk for S-1, especially for HIVAN (Winkler et al. 2010), indicates that the functional causative variant should be extremely rare in the presence of the protective state (G:G) for the S-1

rs5750250 variant. This filtering process yielded the four candidate non-synonymous exonic SNPs for genotyping, shown in Table 2a.

Association analysis: For determining the association of each SNP with ESKD in our dataset, we performed logistic regression with ESKD status (Cases/Control) as the response, and included covariates for local and global ancestry as well as for cohort (African American/Hispanic American). Ancestry estimates were calculated as previously described (Behar et al. 2010). In addition to the four exonic SNPs described in Table 2a, we chose two additional SNPs in the APOL1 region for genotyping, namely SNP rs9622363 located in intron 3, and SNP rs60295735 located in the intergenic region between the genes APOL1 and MYH9. These SNPs were chosen using the same statistical criteria described above which yielded the candidate exonic SNPs.

We tested three major modes of association (recessive, additive, dominant) through definition of appropriate dummy variables in the regression. In addition to this combined analysis, we also performed the regression analysis in each cohort separately, and combined the resulting p-values using Fisher's meta-analysis. Results are shown in Table 1b.

For determining whether the APOL1 SNP rs73885319 explains the association of ESKD with MYH9 SNPs, we performed a logistic regression with the same covariates, but included both APOL1 and MYH9 SNPs in the analysis. To avoid committing to a specific mode of inheritance, we included the SNPs as categorical variables with three values (three possible genotypes), and then performed an analysis of deviance on the results (Hastie and Pregibon 1992), first adding the APOL1 SNP, and then the MYH9 SNP, and performing a chi-square test for the null hypothesis that the MYH9 SNP does not add to our ability to explain ESKD status.

Residual associations of MYH9 SNPs beyond LD with APOL1 missense mutations were performed using analysis of deviance tests (Hastie and Pregibon 1992) of the hypothesis that the APOL1 missense mutation rs73885319 can account for the statistical association between MYH9 SNPs and ESKD status.

## RESULTS AND DICUSSION

We analyzed 119 whole genome sequences recently released by the 1000 Genomes Project (www.1000genomes.org). Of these, 60 are of European origin (CEU), and 59 are of West-African origin (YRI) yielding a total of 7,479 SNPs in the 1.55 Mbp chromosome 22 interval surrounding MYH9 and spanning nucleotide positions 34,000,000 to 35,550,000 (NCBI36). We applied the filtering criteria described in Methods to identify candidates SNPs for further consideration and analysis. Of the 250 variants that met these criteria, which includes all of the six members of the human ApoL gene family, four are coding region non-synonymous mutations (Table 2a). The first two (rs73885319 and rs60910145) are missense mutations in the last exon of the APOL1 gene (S342G and I384M) which is the neighboring gene, located 14 kbp 3' downstream from MYH9. A third SNP (rs11089781) is a nonsense mutation (Q58X) in the APOL3 gene located 110 kbp further 3' downstream. The fourth SNP (rs56767103) is a missense mutation (R71C) in the gene FOXRED2 located 100 kbp upstream to the 5' side of MYH9 (Figure 2). Of note the two variants located 128 bp apart in APOL1 are in almost perfect LD (237 out of 238 chromosomes from the 1000 Genomes Project).

These four variants were genotyped in a previously reported sample set of African American and Hispanic American cases and controls (n=955) (Behar et al. 2010) and association analysis conducted as described in Methods. Table 2a shows the association results for the combined dataset of the African American and Hispanic American cohorts. For comparison, we include the results from the two previously reported leading MYH9 risk variants (S-1 rs5750250, F-1 rs11912763) (Nelson et al. 2010). The APOL1 missense variants (rs73885319 and rs60910145) are more strongly associated with ESKD risk than the leading MYH9 risk variants, both in terms of OR and p values. The lower allele frequency and OR, with a higher p value for the APOL3 nonsense variant, and the weak association for the FOXRED2 missense mutation, render these variants unlikely candidates to explain the risk attributed to this genomic region. We also show that the results for combined and meta-analysis of the two separate cohort-based results are congruent (Table 2b).

Analysis of deviance of the combined logistic regression indicates that LD with APOL1 SNP rs73885319 accounted for much or all the statistical association previously attributed to the leading MYH9 variants with ESKD. In this regard, we also examined two non-coding variants in the APOL1 region which are in high LD with the APOL1 missense mutations, and as expected, both showed significant disease risk association (Figure 3 and Table 2b). Analysis of deviance also shows that the associations of the E-1 and F-1 haplotype SNPs are satisfactorily explained by this mutation ($p>0.5$ for F-1, $p>0.1$ for E-1). For the S-1 SNP rs5750250, we obtain a borderline p-value of 0.01, indicating the possibility that the S-1 SNP may carry an association signal beyond its LD with rs73885319. We expect that this point will be further clarified once additional case control cohorts are examined. Another possibility, is that a one or more additional variants within the region, and in particular among the ApoL gene family members, which did not pass the statistical filter, is also causally related to ESKD risk, such that analysis of deviance based on a single causative variant would not account for all of the MYH9 risk association.

HIVAN has been considered as the most prominent of the non-diabetic forms of kidney disease within what has been termed the MYH9-associated nephropathies (Kopp et al. 2008). We have reported absence of HIVAN in HIV infected Ethiopians, and attributed this to host genomic factors (Behar et al. 2006). Therefore, we examined the allele frequencies of the APOL1 missense mutations in a sample set of 676 individuals from 12 African populations, including 304 individuals from four Ethiopian populations (Table 1). We coupled this with the corresponding distributions for the African ancestry leading MYH9 S-1 and F-1 risk alleles. A pattern of reduced frequency of the APOL1 missense mutations, and also of the MYH9 risk variants was noted in northeastern African in contrast to most central, western and southern African populations examined (Figure 4a). Especially striking was the complete absence of the APOL1 missense mutations in Ethiopia. This combination of the reported lack of HIVAN and observed absence of the APOL1 missense mutations is consistent with APOL1 being functionally relevant gene for HIVAN risk and likely the other forms of kidney disease previously associated with MYH9.

The concomitant absence of the MYH9 F-1 risk variant in the Ethiopian samples examined, is consistent with the causative mutation in APOL1 occurring on a phylogenetic branch in evolution of the region, following the appearance of the S-1 risk and prior to the occurence of the F-1 risk variants, previously reported for MYH9 (Figure 4b). Indeed, F-1 risk homozygotes in our dataset uniformly have the S-1 rs5750250 homozygote risk state (data not shown).

The APOL1 gene encodes apolipoprotein L-1, whose known activities include powerful trypanosomolysis (Lecordier et al. 2009). *Trypanosoma brucei rhodesiense* transmitted by tsetse flies prevalent in central and western Africa is a cause of human African trypanosomiasis, and shows resistance to this trypanosomolytic effect, by virtue of expressing serum resistance–associated protein (SRA), which interacts with the C-terminal domain of apolipoprotein L-1 (Lecordier et al. 2009). The S342G variant is predicted to modify the binding site of the C-terminal domain of the APOL1 gene product (Figure 5a/b/c). With respect to kidney disease risk, APOL1, and other APOL family members are also prominently involved in authophagic pathways (Zhaorigetu et al. 2008), and a recent study has provided compelling evidence for the role of well preserved authophagy in the integrity of renal glomerular podocytes (Hartleben et al. 2010). Moreover, apolipoproteins have also been identified as circulating inhibitors of glomerular proteinuria (Candiano et al. 2001). Functional assays of the effect of the APOL1 missense variants described herein in appropriate experimental model systems will be needed to link the strong and biologically plausible association to a functional pathogenic pathway in kidney disease, and to the possible selective factors which contributed to the observed African allele frequency distribution.

Further studies of variants in other APOL family members are also warranted. The evolution of the APOL gene family in primates, is consistent with their role in protection from pathogens via mechanisms involving programmed cell death of the infected host cell, or alternatively direct pathogen killing as in the case of circulating APOL1 (reference to Smith and Malik). In particular, APOL2, whose kidney expression is restricted to the glomerular compartment, as opposed to APOL1 whose kidney expression is restricted to the tubule compartment (Human Protein Atlas Website link), is warranted. In particular Arg182Cys (rs7285167), which did not pass statistical filtering may still be considered for association and functional studies. Less is known about other non-synonymous variants, such as (Ser155Arg , rs132653 in APOL3) While our current study strongly favors the missense variants Ser342Gly and Ile384Met in the last exon of the APOL1 gene as leading causative candidates, we also raise the possibility that more than one variant among the APOL gene family members may account for disease risk causation and for the MALD peak, which had originally focused attention on MYH9.

The current findings strongly suggest that the intensive efforts (http://www3.niddk.nih.gov/fund/other/MYH9KidneyDisease/) currently underway to identify the ESKD disease phenotype risk causative variant in the chromosome 22 MALD peak should certainly be extended beyond the MYH9 locus. In particular, a strong emphasis should also be placed on the APOL1 missense mutations.

**Acknowledgments**

We thank Steve Smith, Susan Kirby, Rhian Gwilliam, Daniella Magen and Liat Linde for excellent technical assistance. This study was supported by the Canadian and American Technion Societies (Eshagian Estate Fund, Dr. Sidney Kremer Kidney Disease Research Fund) the Israel Science Foundation [890015], and Legacy Heritage Fund to KS and by the European Research Commission [MIRG-CT-2007-208019] and the Israel Science Foundation [1227/09] to SR. DMB thanks the European Commission, Directorate-General for Research for FP7 Ecogene grant 205419.

**Author Contributions**

SR, ST, DMB, SS, and KS conceived the project and planned the experiments. WW conducted clinical characterization and collected samples. NB, AT and EB collected samples. RS, GY, and ST performed genotyping and SNP validation. SR, ST, and KS analyzed and interpreted the data. SR, ST, DMB and KS wrote the paper.

The authors declare no competing financial interests.

# TABLES

**Table 1. The African populations sample set (total n=676), location, and risk allele frequencies for rs73885319 (S342G) in APOL1.**

| Country | Population | Sample Size | Latitude | Longitude | rs73885319 risk allele frequency |
|---|---|---|---|---|---|
| Ghana | Bulsa | 22 | 10.7 | -1.3 | 11% |
| Ghana | Asante | 35 | 5.8 | -2.8 | 41% |
| Cameroon | Somie | 65 | 6.45 | 11.45 | 16% |
| Congo | COG | 55 | -4.25 | 15.28 | 11% |
| Malawi | MWI | 50 | -13.95 | 33.7 | 12% |
| Mozambique | Sena | 51 | -17.45 | 35 | 12% |
| Sudan | Kordofan | 30 | 13.08 | 30.35 | 0% |
| Cameroon | Far-North-CMR/Chad | 64 | 12.5 | 14.5 | 1% |
| Ethiopia | Afar | 76 | 12 | 41.5 | 0% |
| Ethiopia | Amhara | 76 | 11.5 | 38.5 | 0% |
| Ethiopia | Oromo | 76 | 9 | 38.7 | 0% |
| Ethiopia | Maale | 76 | 7.6 | 37.2 | 0% |

**Note to Table 1:** The samples analyzed in Table 1 form part of the collection of DNA maintained by The Centre for Genetic Anthropology at University College London. Buccal cells were collected with informed consent and institutional ethics approval from anonymous donors unrelated at the paternal grandfather level, classified by self declared ethnic identity.

**Table 2a. Association with non-diabetic ESKD of non-synonymous SNPs in APOL1, APOL3 and FOXRED2 in the MALD peak and comparison with leading MYH9 SNPs.**

| rs number | Gene | Type | Chr22 Location[a] | Alleles[b] | YRI risk frequency[c] | CEU risk frequency | Mode[d] | OR | P value |
|---|---|---|---|---|---|---|---|---|---|
| **rs73885319**[e] | **APOL1** | exon 5<br>S342G missense | 34,991,852 | A/**G** | 46% | 0% | **Recessive**<br>**Additive**<br>**Dominant** | **6.7**<br>**2.22**<br>**2.23** | **2.71E-06**<br>**2.38E-08**<br>**8.11E-06** |
| **rs60910145** | **APOL1** | exon 5<br>I384M missense | 34,991,980 | T/**G** | 45% | 0% | **Recessive**<br>**Additive**<br>**Dominant** | **6.74**<br>**2.28**<br>**2.32** | **9.89E-06**<br>**3.00E-08**<br>**4.75E-06** |
| **rs11089781** | **APOL3** | exon 1<br>Q58X nonsense | 34,886,714 | G/**A** | 31% | 0% | Recessive<br>Additive<br>Dominant | 6.62<br>2.18<br>2.22 | 2.82E-03<br>3.79E-06<br>3.23E-05 |
| **rs56767103** | **FOXRED2** | exon 1<br>R71C missense | 35,232,205 | G/**A** | 18% | 0% | Recessive<br>Additive<br>Dominant | 1.33<br>1.52<br>1.66 | 6.83E-01<br>5.19E-02<br>3.64E-02 |
| **rs11912763** | **MYH9** | intron 33<br>F-1 designation | 35,014,668 | G/**A** | 48% | 0% | Recessive<br>Additive<br>Dominant | 2.38<br>1.96<br>2.28 | 2.86E-02<br>4.05E-05<br>4.20E-05 |
| **rs5750250** | **MYH9** | intron13<br>S-1 designation | 35,038,429 | A/**G** | 66% | 6% | Recessive<br>Additive<br>Dominant | 2.48<br>1.78<br>1.55 | 4.29E-05<br>6.68E-05<br>4.97E-02 |

[a] Location on Chromosome 22 in NCBI36 assembly.

[b] African ESKD "risk" state in bold.

[c] Frequencies according to available 1000genome data.

[d] Association results were derived using logistic regression, correcting for global and local African ancestry, and combining the Hispanic and African American cohorts.

[e] See Figure 6 for allele frequency pie-charts in cases versus controls.

## Table 2b. Association of the examined SNPs in the MALD peak with non-diabetic ESKD in African and Hispanic Americans.

| | | | | | | | | Hispanic American | | | | African American | | | | | Combined analysis | |
|---|---|---|---|---|---|---|---|---|---|---|---|---|---|---|---|---|---|---|
| **rs number** | **Chr22 Location** | **Gene** | **Type** | **Alleles** | **YRI risk freq.** | **CEU risk freq.** | **Mode** | **OR** | **lower** | **upper** | **p-value** | **OR** | **lower** | **upper** | **p-value** | **Meta-analysis p-value** | **OR** | **p-value** |
| rs73885319 | 34991852 | APOL1 | exon 5 | A/**G** | 0.457 | 0 | Recessive | 15.48 | 3.99 | 60.00 | 8.8E-04 | 4.86 | 2.35 | 10.06 | 3.5E-04 | 4.9E-06 | 6.70 | 2.7E-06 |
| | | | S342G missense | | | | Additive | 3.59 | 2.21 | 5.83 | 1.5E-05 | 1.90 | 1.46 | 2.48 | 5.9E-05 | 2.0E-08 | 2.22 | 2.4E-08 |
| | | | | | | | Dominant | 3.47 | 1.95 | 6.16 | 3.7E-04 | 1.89 | 1.34 | 2.67 | 2.2E-03 | 1.2E-05 | 2.23 | 8.1E-06 |
| rs60910145 | 34991980 | APOL1 | exon 5 | T/**G** | 0.449 | 0 | Recessive | 12.80 | 3.28 | 49.94 | 2.1E-03 | 5.05 | 2.29 | 11.13 | 7.4E-04 | 2.2E-05 | 6.74 | 9.9E-06 |
| | | | I384M missense | | | | Additive | 3.54 | 2.17 | 5.78 | 2.3E-05 | 1.94 | 1.48 | 2.56 | 7.3E-05 | 3.5E-08 | 2.28 | 3.0E-08 |
| | | | | | | | Dominant | 3.56 | 2.00 | 6.33 | 3.0E-04 | 1.95 | 1.37 | 2.76 | 1.8E-03 | 8.1E-06 | 2.32 | 4.8E-06 |
| rs60295735 | 34997100 | Intergenic | Intergenic | G/**A** | 0.432 | 0 | Recessive | 12.79 | 3.28 | 49.92 | 2.1E-03 | 2.99 | 1.56 | 5.73 | 5.8E-03 | 1.5E-04 | 4.27 | 1.2E-04 |
| | | | (APOL1-MYH9) | | | | Additive | 3.43 | 2.10 | 5.60 | 3.6E-05 | 1.76 | 1.35 | 2.31 | 5.6E-04 | 3.8E-07 | 2.10 | 4.6E-07 |
| | | | | | | | Dominant | 3.32 | 1.87 | 5.91 | 5.9E-04 | 1.87 | 1.31 | 2.66 | 3.5E-03 | 2.9E-05 | 2.22 | 1.5E-05 |
| rs9622363 | 34986501 | APOL1 | intronic | A/**G** | 0.711 | 0 | Recessive | 5.11 | 2.36 | 11.06 | 5.2E-04 | 3.68 | 2.46 | 5.52 | 1.2E-07 | 1.5E-09 | 3.92 | 6.3E-10 |
| | | | | | | | Additive | 2.80 | 1.73 | 4.53 | 4.2E-04 | 2.34 | 1.79 | 3.06 | 1.8E-07 | 1.8E-09 | 2.46 | 2.6E-10 |
| | | | | | | | Dominant | 2.03 | 1.15 | 3.60 | 4.1E-02 | 2.26 | 1.41 | 3.61 | 4.3E-03 | 1.7E-03 | 2.20 | 3.2E-04 |
| rs56767103 | 35232205 | FOXRED2 | exon 1 | G/**A** | 0.177 | 0 | Recessive | Inf | 0.00 | Inf | 9.9E-01 | 1.02 | 0.33 | 3.17 | 9.8E-01 | 1.0E+00 | 1.33 | 6.8E-01 |
| | | | R71C missense | | | | Additive | 2.75 | 1.40 | 5.41 | 1.4E-02 | 1.23 | 0.84 | 1.82 | 3.7E-01 | 3.3E-02 | 1.52 | 5.2E-02 |
| | | | | | | | Dominant | 2.69 | 1.33 | 5.46 | 2.1E-02 | 1.33 | 0.85 | 2.10 | 3.0E-01 | 3.9E-02 | 1.66 | 3.6E-02 |
| rs11089781 | 34886714 | APOL3 | exon 1 | G/**A** | 0.305 | 0 | Recessive | 2.30 | 0.42 | 12.44 | 4.2E-01 | 13.06 | 2.43 | 70.14 | 1.2E-02 | 3.1E-02 | 6.62 | 2.8E-03 |
| | | | Q58X nonsense | | | | Additive | 2.57 | 1.55 | 4.24 | 2.0E-03 | 2.01 | 1.45 | 2.78 | 4.2E-04 | 1.3E-05 | 2.18 | 3.8E-06 |
| | | | | | | | Dominant | 2.87 | 1.66 | 4.96 | 1.5E-03 | 1.93 | 1.32 | 2.82 | 4.3E-03 | 8.2E-05 | 2.22 | 3.2E-05 |
| rs4821480 | 35025193 | MYH9 | intron23 | T/**G** | 0.763 | 0.058 | Recessive | 3.37 | 1.56 | 7.29 | 9.6E-03 | 1.82 | 1.24 | 2.69 | 1.1E-02 | 1.1E-03 | 2.05 | 6.6E-04 |
| | | | E-1 designation | | | | Additive | 1.67 | 1.06 | 2.63 | 6.4E-02 | 1.64 | 1.23 | 2.19 | 4.4E-03 | 2.6E-03 | 1.65 | 6.5E-04 |
| | | | | | | | Dominant | 1.14 | 0.66 | 1.99 | 6.9E-01 | 1.92 | 1.10 | 3.36 | 5.6E-02 | 1.6E-01 | 1.47 | 1.0E-01 |
| rs5750250 | 35038429 | MYH9 | intron13 | A/**G** | 0.661 | 0.058 | Recessive | 3.82 | 1.67 | 8.73 | 7.5E-03 | 2.29 | 1.54 | 3.43 | 6.7E-04 | 6.7E-05 | 2.48 | 4.3E-05 |
| | | | S-1 designation | | | | Additive | 1.50 | 0.96 | 2.34 | 1.4E-01 | 1.92 | 1.45 | 2.54 | 1.3E-04 | 2.1E-04 | 1.78 | 6.7E-05 |
| | | | | | | | Dominant | 1.02 | 0.60 | 1.74 | 9.5E-01 | 2.28 | 1.37 | 3.81 | 7.8E-03 | 4.4E-02 | 1.55 | 5.0E-02 |
| rs11912763 | 35014668 | MYH9 | intron 33 | G/**A** | 0.483 | 0 | Recessive | 4.31 | 1.21 | 15.34 | 5.9E-02 | 1.95 | 0.95 | 3.99 | 1.3E-01 | 4.4E-02 | 2.38 | 2.9E-02 |
| | | | F-1 designation | | | | Additive | 3.02 | 1.80 | 5.08 | 4.7E-04 | 1.67 | 1.23 | 2.27 | 5.8E-03 | 3.8E-05 | 1.96 | 4.1E-05 |
| | | | | | | | Dominant | 3.64 | 1.97 | 6.75 | 5.7E-04 | 1.90 | 1.29 | 2.79 | 6.2E-03 | 4.8E-05 | 2.28 | 4.2E-05 |

**Note to Table 2b:** Table 2b includes in addition to the SNPs in Table 2a, also similarly derived results on several previously described associated SNPs in the MYH9 gene. The Table also demonstrates similarity between the p values obtained from combining the p values from the separate cohort based analyses (African American, Hispanic American) in a meta analysis, and the p values obtained directly from a combined analysis of both cohorts including an indicator for cohort. This demonstrates the robustness of the statistical conclusions to variations in methodology.

## FIGURES

**Figure 1. RFLP reaction for APOL1 missense mutations.**

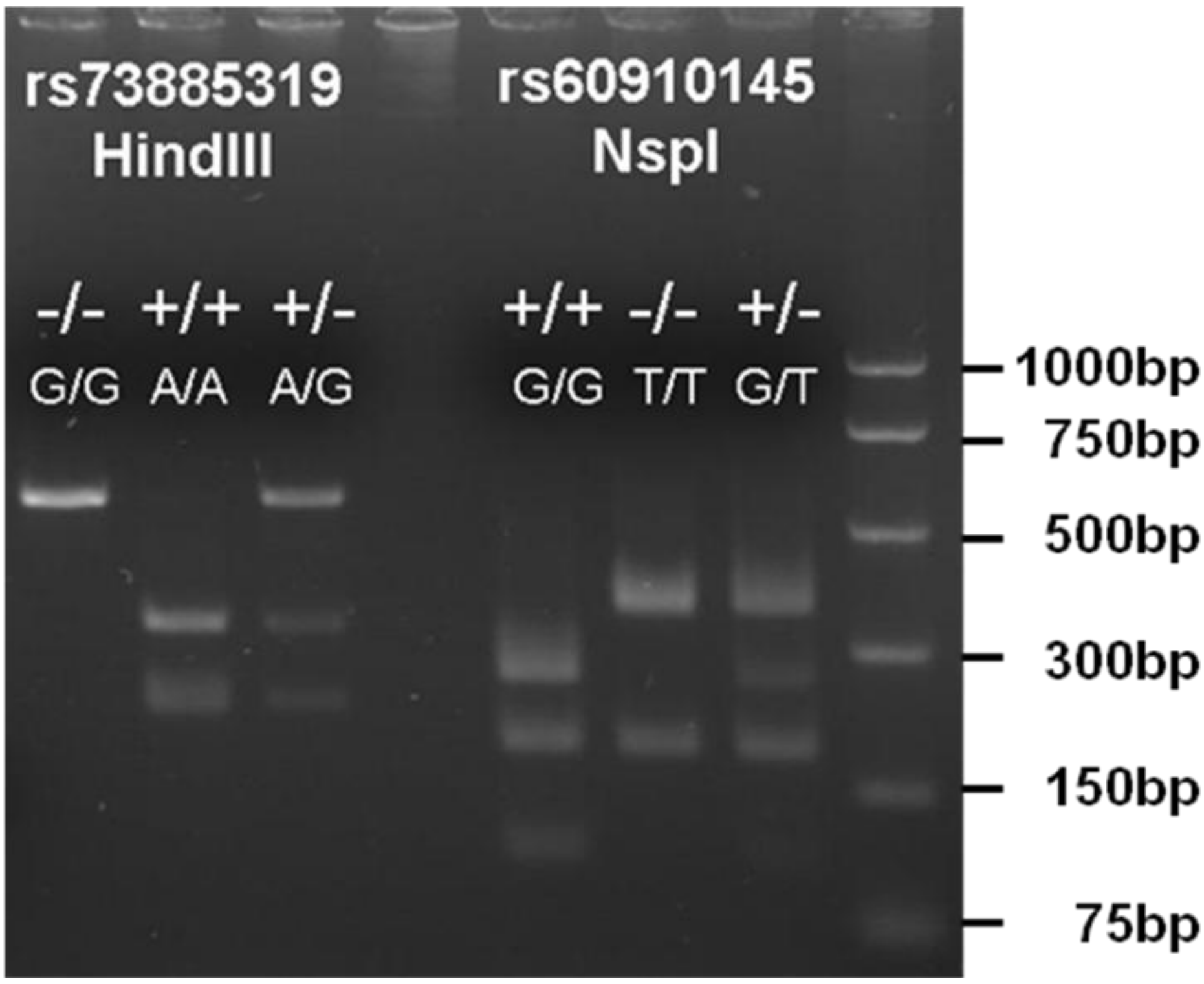

SNP RS73885319 A/G: The allele containing G eliminates APOL1 SNP rs73885319 (alleles A/G) eliminated a recognition site of endonuclease Hind III. SNP RS60910145 T/G: The allele containing TAPOL1 SNP rs60910145 (alleles T/G) generates a new recognition site of endonuclease NspI. Thus to confirm and screen for the mutations, we amplified by PCR a 538 bp fragment that contains both these SNPs. For PCR amplification we used the forward primer: 5'- ACA AGC CCA AGC CCA CGA CC-3' and the reverse primer: 5'- CCT GGC CCC TGC CAG GCA TA-3'. PCR reaction mix included 30ng template DNA with 7.5 pmol of each primer and we used the Red Load Taq Master (Larova). PCR conditions were 95°c for 3 minutes followed by 40 cycles at 95°c for 30 seconds, 65°c for 20 seconds, 72°c for 1 minute ; the resulting amplicon was digested with both endonucleases Hind III and NspI. The resulting amplicon was digested separately with endonucleases Hind III and NspI, and run on a 2% agarose gel.

**Figure 2. Schematic view of the chromosomal region encompassing the examined SNPs.**

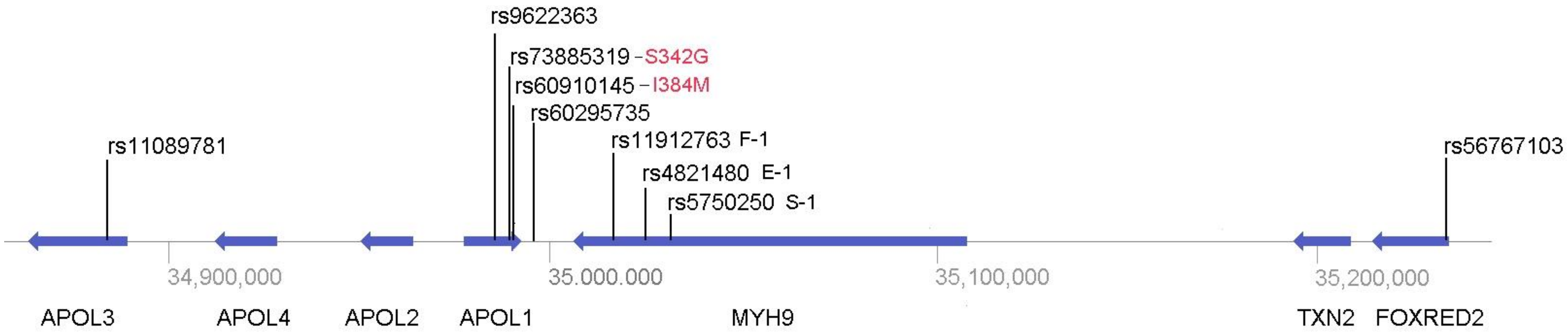

**Figure 3. LD plot of the ESKD associated SNPs in the APOL1 and MYH9 region.**

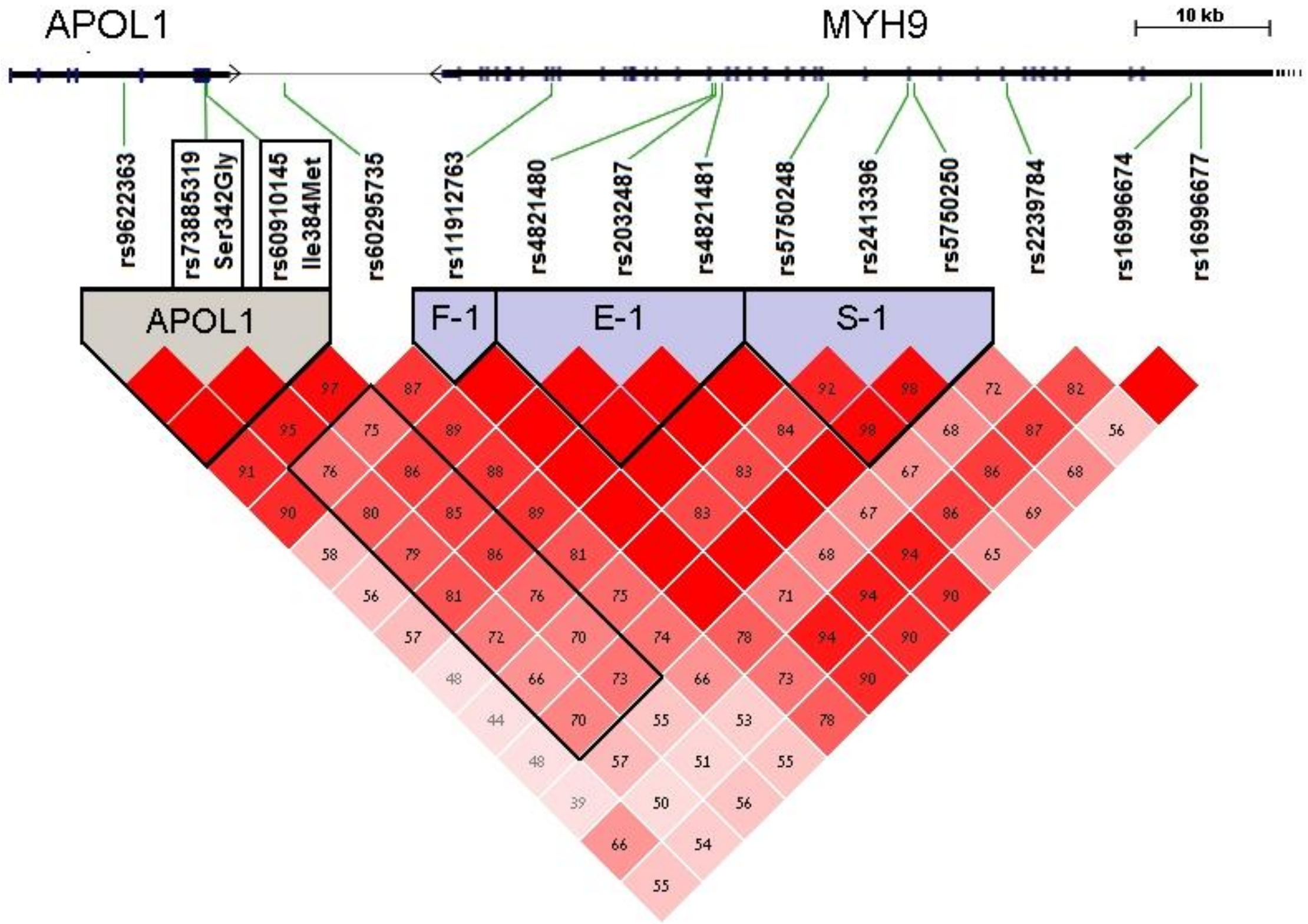

Linkage disequilibrium (LD) plot of non-diabetic ESKD associated SNPs in APOL1 and MYH9 genes with their physical locations on chromosome 22. The color scheme represents the pairwise linkage disequilibrium value (D'/LOD) for the 4 new SNPs outside MYH9 (2 of which are missense mutations in APOL1) and for the previously published 10 MYH9 SNPs described in (Behar et al. 2010). The LD plot was calculated based on the African American control samples (n = 140). The plot was generated using the program HaploView (Barrett et al. 2005). Bright red squares presents SNPs with linkage LOD ≥ 2 and D'=1.

**Figure 4a. Spatial allele frequency distributions in Africa of the ESKD risk variants.**

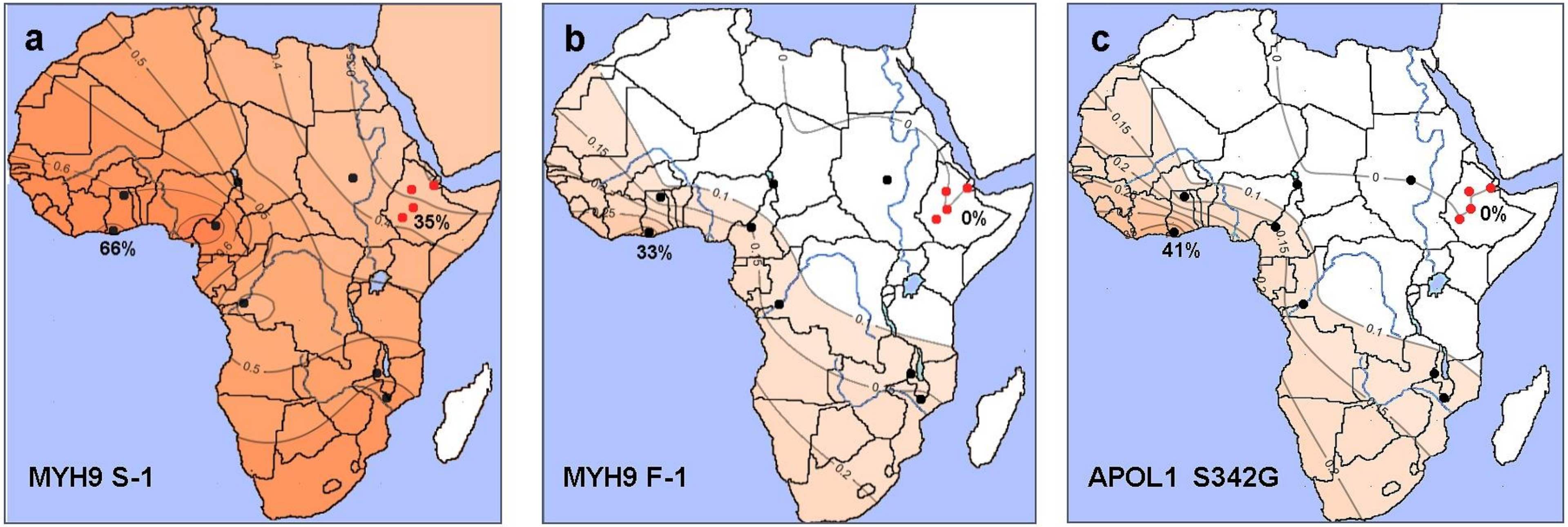

The given contour maps correspond to: a) MYH9 S-1 SNP rs5750250); b) MYH9 F-1 SNP (rs11912763); c) APOL1 S342G missense mutation (rs73885319). Maps were generated based on genotyping 12 African populations (n=676) (Table 1), using Surfer V.9 (Golden Software). Populations locations are marked (red circles for Ethiopia). Risk allele frequencies in Ethiopia and in South-Ghana are indicated. Notably, the most strongly associated MYH9 S-1 rs5750250 risk variant retains a 35% allele frequency in Ethiopians, while the Ethiopian allele frequencies for MYH9 F-1 rs11912763 and APOL1 missense mutation rs73885319 are zero. This pattern is consistent with the occurrence of the APOL1 missense mutations on a phylogenetic branch of the genomic region following the appearance of the MYH9 S-1 risk variants but prior to the appearance of the MYH9 F-1 risk variant. Indeed, F-1 risk homozygotes in our dataset uniformly have the S-1 rs5750250 homozygote risk state (data not shown).

**Figure 4b. A phylogenetic tree of the MYH9 and APOL1 phased haplotypes.**

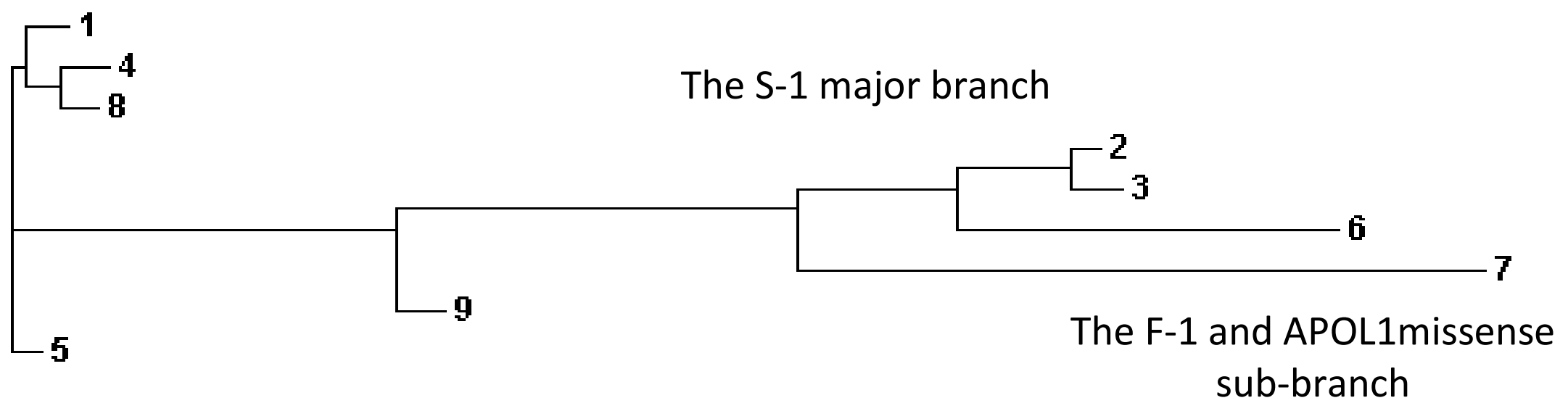

Note: this tree was created by the program PHASE, based on 28 SNPs across the genes MYH9 and APOL1.

**Figures 5 (a,b and c). Predicted peptide structures of the C-terminus domain of the APOL1 gene product (amino acid positions 339-398) that contains the missense mutations S342G and I384M.** All predictions were generated using the program I-TASSER (Zhang 2008, 2009), structures were edited with the program CHIMERA (Pettersen et al. 2004). All suggested predicted structures had Tm value >0.5.

a) Predicted structure and location of amino acid changes. C-terminus domain is predicted to have a bent alpha-helix structure. The mutation I384M is located on the external surface of the predicted alpha-helix, while the S342G is buried inside.

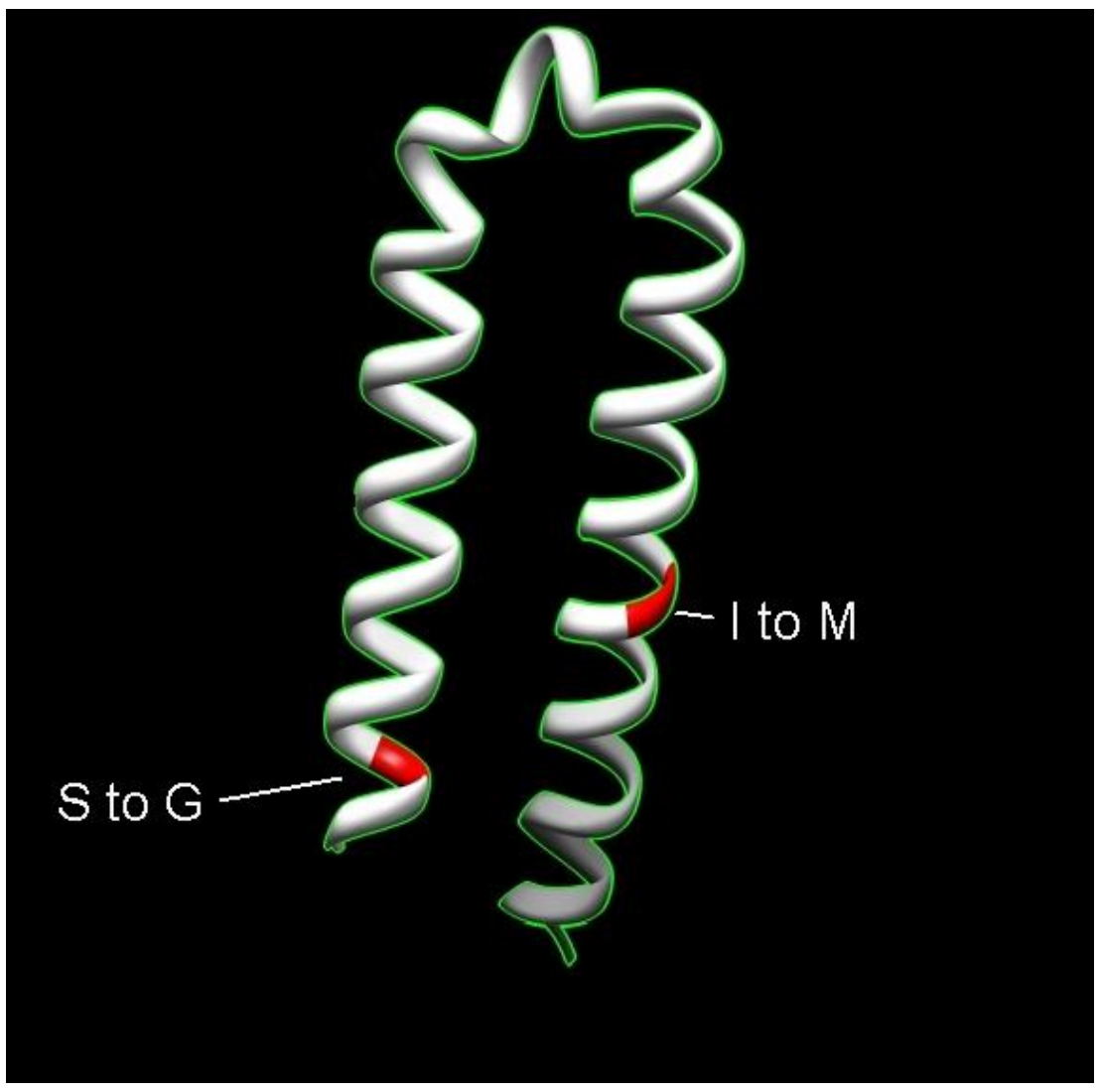

b) Hydrophobicity of the predicted peptide surface (RED- hydrophobic amino-acids, BLUE- polar amino-acids). Hydrophobic core is predicted to stabilize the bent C-terminus helical structure.

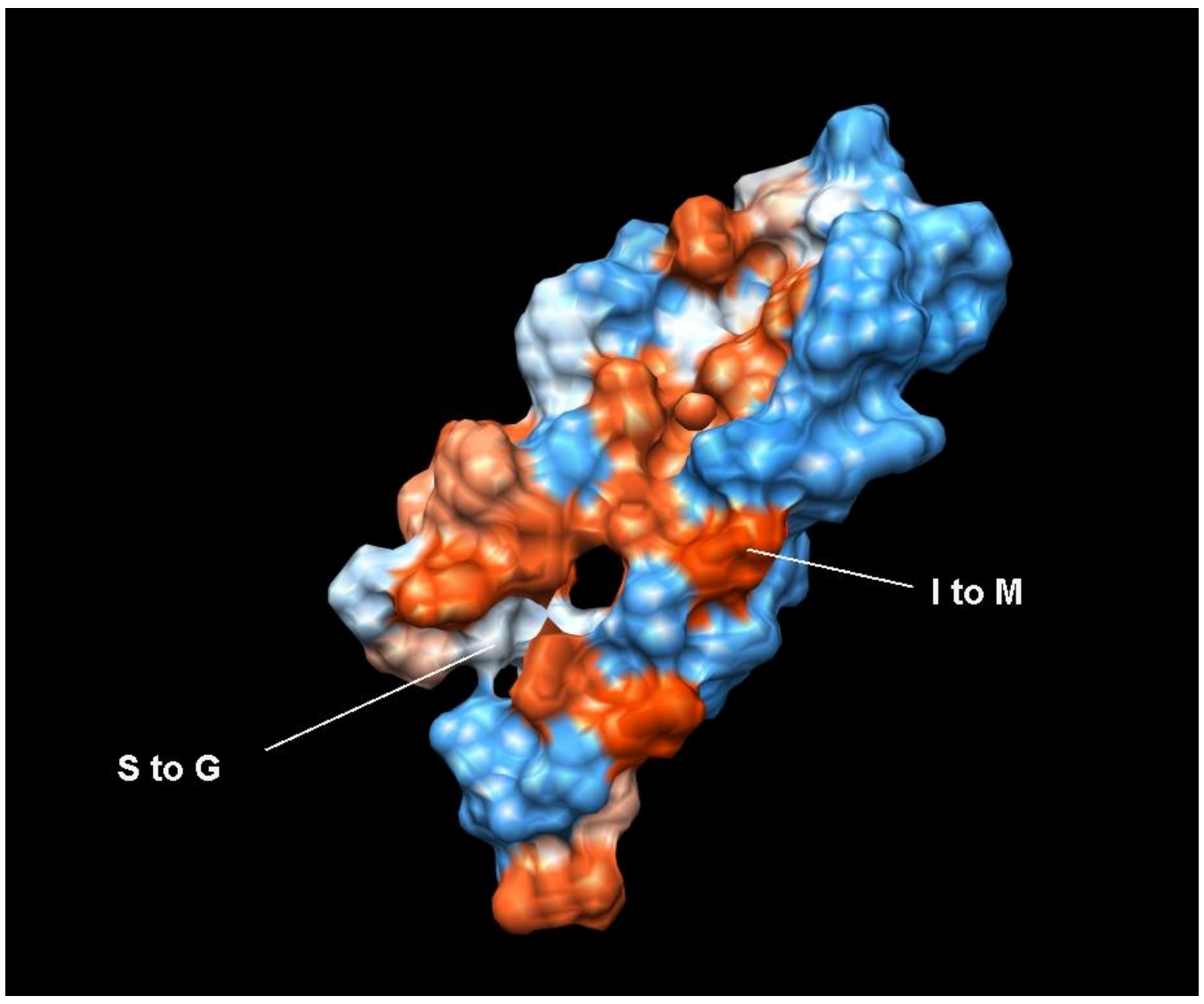

c) Identified binding site in the predicted structure of the APOL1 C-terminus domain (Zhang 2008, 2009), based on similarity to an analogous known binding site (Billas et al. 2003). S342G is involved in the predicted binding site domain, and is predicted to modify its binding ability.

:

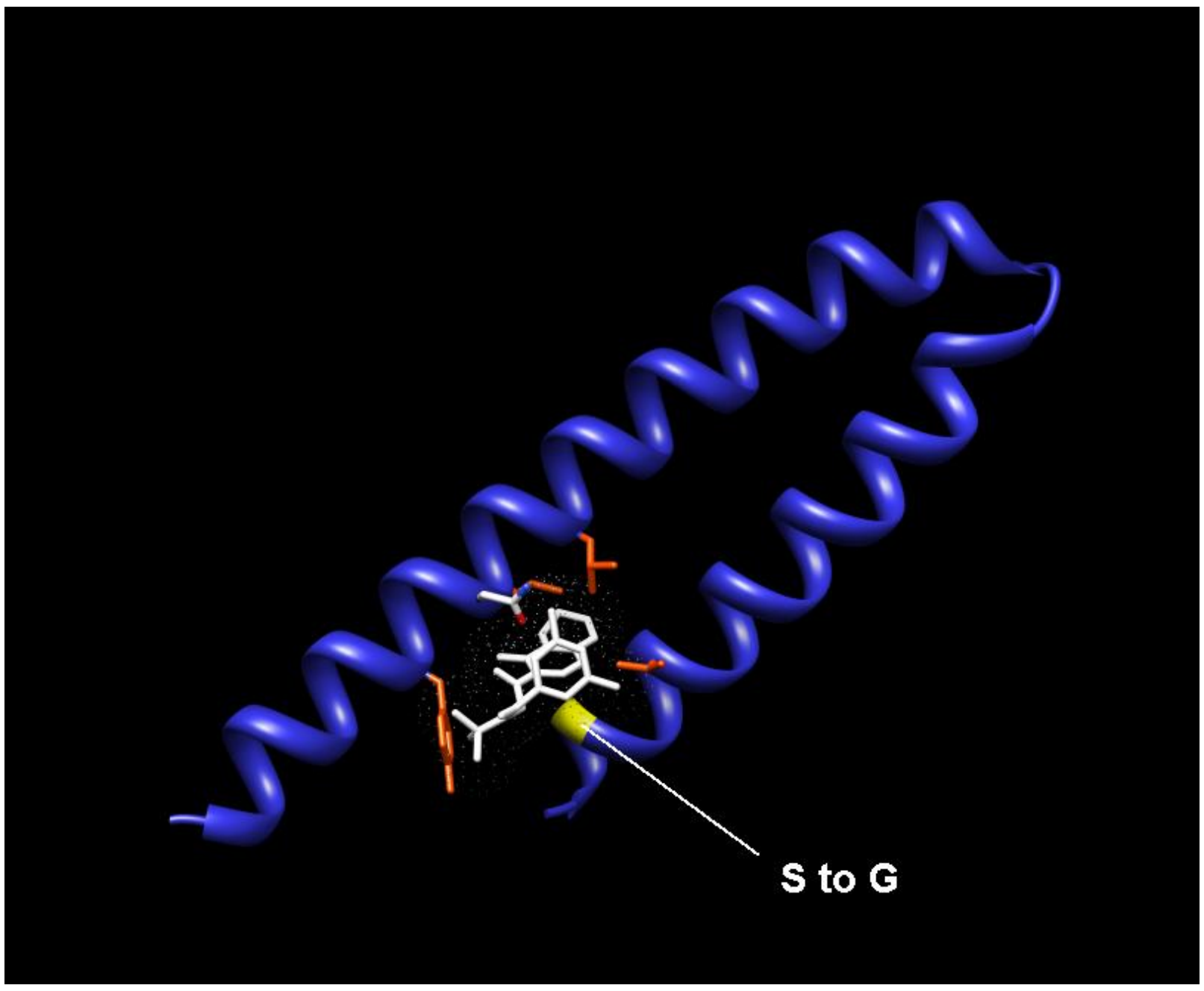

**Figure 6. Pie charts of allele frequencies for the APOL1 SNP rs73885319 (S342G) in African Americans and Hispanic Americans cases versus controls.**

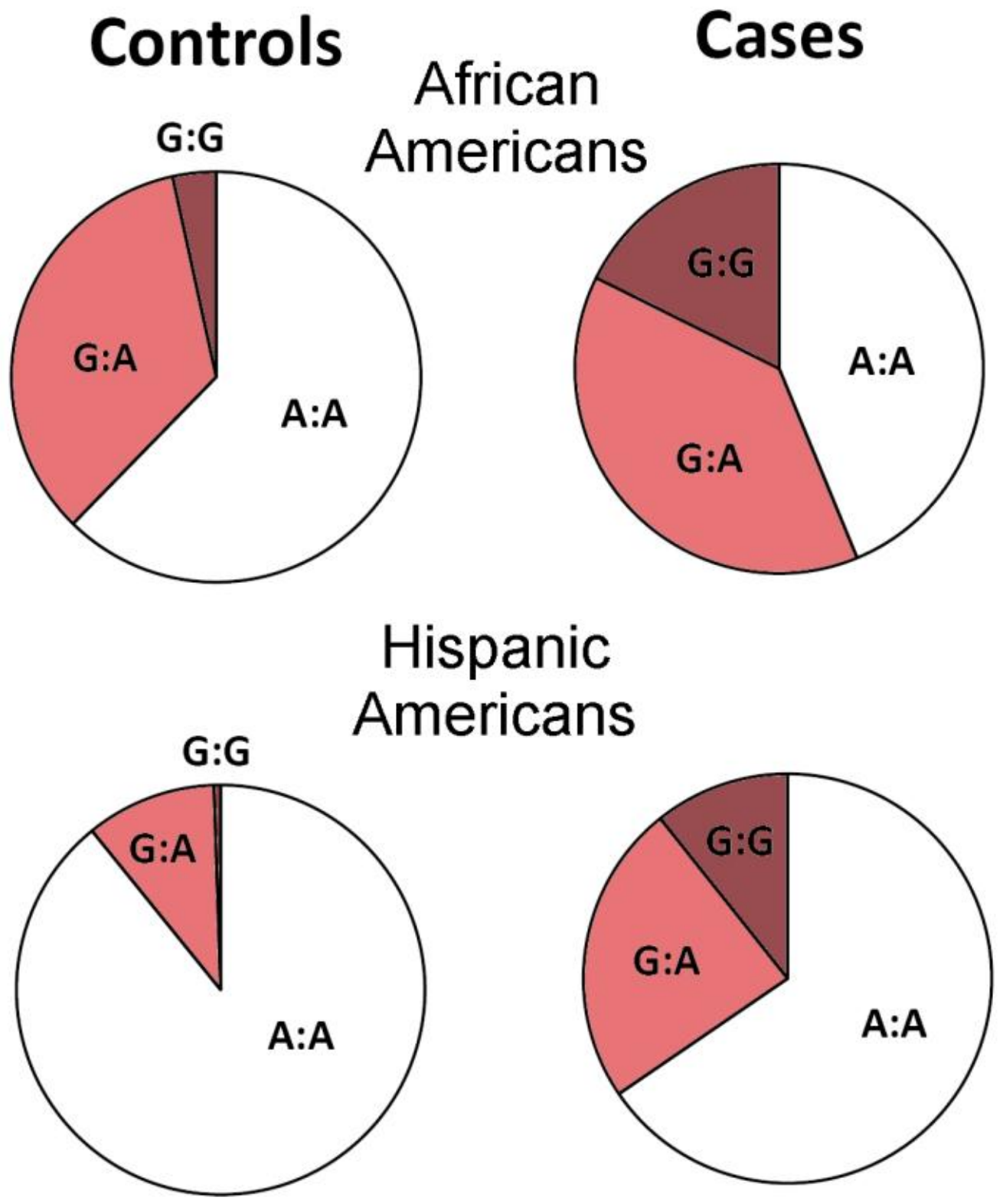

References

Behar DM, Rosset S, Tzur S, Selig S, Yudkovsky G, Bercovici S, Kopp JB, Winkler CA, Nelson GW, Wasser WG, Skorecki K (2010) African ancestry allelic variation at the MYH9 gene contributes to increased susceptibility to non-diabetic end-stage kidney disease in Hispanic Americans. Hum Mol Genet 19: 1816-27

Behar DM, Shlush LI, Maor C, Lorber M, Skorecki K (2006) Absence of HIV-associated nephropathy in Ethiopians. Am J Kidney Dis 47: 88-94

Structural (Billas IM, Iwema T, Garnier JM, Mitschler A, Rochel N, Moras D (2003 adaptability in the ligand-binding pocket of the ecdysone hormone receptor. Nature 426: 91-6

Bostrom MA, Freedman BI (2010) The Spectrum of MYH9-Associated Nephropathy. Clin J Am Soc Nephrol

Candiano G, Musante L, Carraro M, Faccini L, Campanacci L, Zennaro C, Artero M, Ginevri F, Perfumo F, Gusmano R, Ghiggeri GM (2001) Apolipoproteins prevent glomerular albumin permeability induced in vitro by serum from patients with focal segmental glomerulosclerosis. J Am Soc Nephrol 12: 143-50

Hartleben B, Godel M, Meyer-Schwesinger C, Liu S, Ulrich T, Kobler S, Wiech T, Grahammer F, Arnold SJ, Lindenmeyer MT, Cohen CD, Pavenstadt H, Kerjaschki D, Mizushima N, Shaw AS, Walz G, Huber TB (2010) Autophagy influences glomerular disease susceptibility and maintains podocyte homeostasis in aging mice. J Clin Invest 120: 1084-96

Hastie TJ, Pregibon D (1992) In: Chambers JM, Hastie TJ (eds) Statistical Models in S. Wadsworth & Brooks

Kao WH, Klag MJ, Meoni LA, Reich D, Berthier-Schaad Y, Li M, Coresh J, Patterson N, Tandon A, Powe NR, Fink NE, Sadler JH, Weir MR, Abboud HE, Adler SG, Divers J, Iyengar SK, Freedman BI, Kimmel PL, Knowler WC, Kohn OF, Kramp K, Leehey DJ, Nicholas SB, Pahl MV, Schelling JR, Sedor RS (2008) MYH9 is JR, Thornley-Brown D, Winkler CA, Smith MW, Parekh associated with nondiabetic end-stage renal disease in African Americans. Nat Genet 40: 1185-92

Kopp JB, Smith MW, Nelson GW, Johnson RC, Freedman BI, Bowden DW, Oleksyk Dart ,T, McKenzie LM, Kajiyama H, Ahuja TS, Berns JS, Briggs W, Cho ME RA, Kimmel PL, Korbet SM, Michel DM, Mokrzycki MH, Schelling JR, Simon E, Trachtman H, Vlahov D, Winkler CA (2008) MYH9 is a major-effect risk gene for focal segmental glomerulosclerosis. Nat Genet 40: 1175-84

P, Tebabi P, Paturiaux-Hanocq F, Andris F, Lecordier L, Vanhollebeke B, Poelvoorde Lins L, Pays E (2009) C-terminal mutants of apolipoprotein L-I efficiently kill both Trypanosoma brucei brucei and Trypanosoma brucei rhodesiense. PLoS Pathog 5: e1000685

Nelson GW, Freedman BI, Bowden DW, Langefeld CD, An P, Hicks PJ, Bostrom MA, Johnson RC, Kopp JB, Winkler CA (2010) Dense mapping of MYH9 localizes the strongest kidney disease associations to the region of introns 13 to 15. Hum Mol Genet 19: 1805-15

Petkov PM, Ding Y, Cassell MA, Zhang W, Wagner G, Sargent EE, Asquith S, Crew V, Johnson KA, Robinson P, Scott VE, Wiles MV (2004) An efficient SNP system for mouse genome scanning and elucidating strain relationships. Genome Res 14: 1806-11

Pettersen EF, Goddard TD, Huang CC, Couch GS, Greenblatt DM, Meng EC, Ferrin TE (2004) UCSF Chimera--a visualization system for exploratory research and analysis. J Comput Chem 25: 1605-12 van Eijk R, van Puijenbroek M, Chhatta AR, Gupta N, Vossen RH, Lips EH, Cleton-Jansen AM, Morreau H, van Wezel T (2010) Sensitive and specific KRAS somatic mutation analysis on whole-genome amplified DNA from archival tissues. J Mol Diagn 12: 27-34

Winkler CA, Nelson G, Oleksyk TK, Nava MB, Kopp JB (2010) Genetics of focal segmental glomerulosclerosis and human immunodeficiency virus-associated collapsing glomerulopathy: the role of MYH9 genetic variation. Semin Nephrol 30: 111-25

Zhang Y (2008) I-TASSER server for protein 3D structure prediction. BMC Bioinformatics 9: 40

Zhang Y (2009) I-TASSER: fully automated protein structure prediction in CASP8. Proteins 77 Suppl 9: 100-13

Zhaorigetu S, Wan G, Kaini R, Jiang Z, Hu CA (2008) ApoL1, a BH3-only lipid-binding protein, induces autophagic cell death. Autophagy 4: 1079-82